 \documentclass[12pt,preprint]{aastex}
\newcommand       \Msun        	{$M_{\odot}$}
\newcommand       \Lsun      	{$L_{\odot}$} 
\newcommand	     \Mpc              {Mpc$^{-3}$}
\newcommand       \mum        	 {$\mu$m}
\newcommand       \nun		{$\nu$}
\newcommand        \lya		{Ly$\alpha$}
\newcommand        \ha		{H$\alpha$}
\newcommand	     \zmin      	{$z_{min}$}
\newcommand      \zmax      	{$z_{max}$}
\newcommand 	      \nwats            {nW m$^{-2}$ sr$^{-1}$}
\newcommand       \yr                 {yr$^{-1}$}
\newcommand      \gray       {$\gamma$--ray}
\begin{document}

\title{THE NEAR INFRARED BACKGROUND: \\ INTERPLANETARY DUST OR PRIMORDIAL  STARS?}

\author{Eli Dwek\altaffilmark{1}, Richard G. Arendt\altaffilmark{2}, and Frank Krennrich\altaffilmark{3}, 
}
\altaffiltext{1}{Observational Cosmology Laboratory, Code 665, NASA Goddard Space Flight Center,
Greenbelt, MD 20771, e-mail: eli.dwek@nasa.gov}
\altaffiltext{2}{Science Systems and Applications, Inc., Observational Cosmology Laboratory, Code 665, NASA Goddard Space Flight Center, Greenbelt, MD 20771, e-mail: arendt@milkyway.gsfc.nasa.gov}
\altaffiltext{3}{Department of Physics and Astronomy, Iowa State University, Ames, IA 50011, e-mail: krennich@iastate.edu}

\begin{abstract}
The intensity of the diffuse $\sim$ 1 - 4 \mum\ sky emission from which solar system and Galactic foregrounds have been subtracted is in excess of that expected from energy released by galaxies and stars that formed during the $z \lesssim 5$ redshift interval (Arendt \& Dwek 2003, Matsumoto et al. 2005). The spectral signature of this excess near-infrared background light (NIRBL) component is almost identical to that of reflected sunlight from the interplanetary dust cloud, and could therefore be the result of the incomplete subtraction of this foreground emission component from the diffuse sky maps. Alternatively, this emission component could be extragalactic. Its spectral signature is consistent with that of redshifted continuum and recombination line emission from H~II regions formed by the first generation of very massive stars. In this paper we analyze the implications of this spectral component for the formation rate of these Population~III (Pop~III) stars, the redshift interval during which they formed, the reionization of the universe and evolution of collapsed halo masses. Assuming that these Pop~III stars are massive objects radiating at the Eddington luminosity and ending their lives by directly collapsing into black holes, we find that to reproduce the intensity and spectral shape of the NIRBL requires a  peak star formation rate of $\sim$2.5~\Msun\ yr$^{-1}$~\Mpc, with a $(1+z)^{-2}$ dependence on redshift, until the epoch ended at redshifts $z \approx$ 7--9.  It requires a comoving luminosity density of about 2.7$\times10^{11}$~\Lsun\ Mpc$^{-3}$ corresponding to a total energy input of 670--820~keV per baryon, and that about 10\% of the total number of baryons in the universe be converted to Pop~III stars. All these numbers are higher by about a factor of 4 to 10 than those derived from models in which  Pop~III stars form at a rate that is proportional to the collapse rate of haloes in a cold dark matter dominated universe. Furthermore, an extragalactic origin for the NIRBL leads to physically unrealistic absorption-corrected spectra of distant TeV blazars. All these results suggest that Pop~III stars contribute only a fraction of the NIRBL intensity with zodiacal light, star forming galaxies, and/or non-nuclear sources giving rise to the remaining fraction.
Further 0.1 to 10 \mum\ observations of the diffuse sky and the zodiacal cloud are therefore crucial for resolving the true spectrum and origin of the NIRBL.
 \end{abstract}

\section{INTRODUCTION}

The extragalactic background light (EBL) consists of all radiative energy outputs, whether powered by nuclear or gravitational processes, that were released into the universe after the epoch of recombination. At UV to near-infrared wavelengths it comprises the fraction of that radiation that was not absorbed by dust and reradiated at  mid- to far-infrared (IR) wavelengths. In addition to the radiative output from star forming galaxies and active galactic nuclei, the EBL can also harbor the radiative imprint of a variety of "exotic" objects including exploding stars, primordial black holes, decaying particles, and primordial very massive objects (Bond, Carr, \& Hogan 1991). Assuming that all the dark matter required to close the universe was contained in these objects, Bond, Carr, \& Hogan (1991) suggested that they could potentially have an important effect on the spectrum of the EBL. However, subsequent limits on and detection of the EBL ( Hauser et al. 1998; Hauser \& Dwek 2001), strongly constrained the mass of some of these objects. Furthermore, the lack of any strong physical motivation for their existence rendered their contribution highly speculative. 

Quasars offer an alternative method for probing the formation and evolution of the first collapsed objects in the universe. The absorption spectra of high redshift quasars can reveal the ionization state of the intergalactic medium (IGM), and therefore provide direct  evidence for the presence or absence of ionizing sources at high redshifts. The absence of a Gunn-Peterson trough in the spectrum of the luminous quasar SDSS~1044-0125  located at $z$ = 5.8 suggested that universe was already  highly ionized at that redshift (Fan et al. 2000). The detection of excess power in the polarization-temperature cross-power spectrum of the cosmic microwave background with the {\it Wilkinson Microwave Anisotropy Probe} ({\it WMAP}) at large angular scales confirmed that the universe was reionized at high redshifts  (Bennett et al. 2003). Assuming that the reionization was instantaneous and complete, Kogut et al. (2003) derived an optical depth of $\tau \approx 0.17 \pm 0.4$ to reionization, corresponding to a redshift of $z = 17 \pm 3$. However, the earlier detection of Gunn-Peterson trough in the spectra of several quasars with redshifts between $\sim$ 5.8 and 6.3 (Fan et al. 2001) suggests that neutral hydrogen was present at those redshifts in sufficient quantities to lead to the absorption of all photons shortward of the Lyman limits from their spectra. These seemingly conflicting results suggest that the reionization of the universe did not occur in a single event, rendering the formation history and evolution of the first objects capable of producing ionizing radiation a question of great cosmological importance.

The formation of these first ionizing stars is relatively well understood since it took place under conditions that are considerably simpler than those governing present day star formation: a well defined set of initial conditions needed to calculate the growth of dark matter density perturbations, a metal-free gas simplifying calculations of the chemistry and cooling of the collapsing baryonic matter, and the absence of magnetic fields (see recent reviews by Loeb \& Barkana 2001, Barkana \& Loeb 2001, Bromm \& Larson 2004, Glover 2004, and Kashlinsky 2005). In spite of this relative simplicity, there are still many unanswered questions regarding the formation and nature of these first stellar objects:  (1) when did these objects first form, and when did they stop forming?; (2) what is the fraction of the collapsing gas that actually formed stars?; (3) how did the gas fragment and what is the stellar initial mass function?; (4) how did the ionizing stellar radiation interact with the ambient gas, and how did these primordial H~II regions evolve? 

Searches for answers to these questions have led to a flurry of models and predictions of various observational effects associated with these Population~III (Pop~III) stars, such as their effect on the early helium and metal enrichment of the universe (e.g. Haiman \& Loeb 1997; Salvaterra \& Ferrara 2003b; Scannapieco, Schneider, \& Ferrara 2003), or their anisotropies (Kashlinsky et al. 2004, Cooray et al. 2004, Magliocchetti et al. 2003). In particular, Santos, Bromm, \& Kamionkowski (2002, SBK), Cooray \& Yoshida (2004), and Madau \& Silk (2005) reconsidered their potential imprint on the spectrum of the EBL. More recently,  Salvaterra \& Ferrara (2003a) claimed the actual detection of their signature in the $\sim$ 1 to 4~\mum\ wavelength region of the EBL.

The claimed detection of the Pop~III signature in  the EBL relies on the  {\it COBE}/DIRBE (Hauser et al. 1998) and {\it IRTS}/NIRS (Matsumoto et al. 2000, 2005) 1 - 4~\mum\ measurements, assuming their extragalactic origin. The DIRBE and NIRS residuals, obtained after the subtraction of local foreground emission components from the diffuse sky, are higher than the ground-based measurements of the integrated light from resolved galaxies (Madau \& Pozzetti 2000) or even their extrapolated intensity based on the evolutionary models  (Totani et al. 2001). The {\it HST} and ground-based optical detections of the EBL by Bernstein, Freedman, \& Madore (2002) are also in excess of their ground-based counterparts, but by a smaller amount. 
The near-IR ($\sim$ 1 to 4~\mum) region of the EBL may therefore be a distinct spectral component which may be either of local or extragalactic origin.

In this paper we first describe the data and examine the nature of the excess emission component, whether it is truly extragalactic or whether it reflects systematic errors in the subtraction of foreground emission from the zodiacal dust cloud (\S2). Assuming its extragalactic nature, we fit this component with the contribution of the line and continuum emission from metal-free Pop~III stars and their surrounding nebulae (\S3). Our approach differs from that of SBK or  Cooray \& Yoshida (2004) who calculated the EBL expected from Pop~III stars forming at a predetermined rate inferred from the collapse and merging rate of dark matter halos. These studies did not attempt to fit the DIRBE or NIRS data, and consequently, their model predictions fall short of the excess $\sim$ 1 - 4 \mum\ emission. In contrast, Salvaterra \& Ferrara (2003a) and Madau \& Silk (2005) assumed that the excess emission is extragalactic, and explored the possibility that it was formed by Pop~III stars. However, Salvaterra \& Ferrara (2003a) failed to explore the implications of their model for the formation rate of these objects. Madau \& Silk (2005) assumed all the stellar energy to be emitted as Ly$\alpha$ radiation, and adopted a {\it very} conservative estimate for the 1.25 \mum\ NIRBL intensity of only 2.5 \nwats, instead of the value of $\sim$ 70~\nwats\ implied by the observed excess and fitted by Salvaterra \& Ferrara (2003a). 

In our model we assume that the Pop~III stars form an ionization-bounded H~II region, so that all their ionizing photons are absorbed in the remaining nebular gas from which they were formed instead of escaping into the intergalactic medium (IGM), and use CLOUDY to calculate the emerging stellar and nebular emission.
Assuming a continuous star formation process over a given redhift interval, we calculate the spectral signature of these objects for a variety of model parameters. Fitting these spectra to the observations we derive the epoch over which Pop~III stars formed, their formation rate, and the total energy released during their lifetime. Cosmological implications  are presented in \S4. In \S5 we briefly describe other considerations that can shed light on the origin of the EBL, namely, fluctuations in the EBL and the effect of the NIRBL on the absorption-corrected spectra of blazars. The results of our paper are briefly summarized in \S6. 

In all our calculations we adopt a cosmological model with parameters determined from the analysis of the ({\it WMAP}, Bennett et al. 2003): a dark energy density of $\Omega_{\Lambda}$ = 0.73,  and total and baryonic matter densities of $\Omega_m$ = 0.27 and $\Omega_b$ = 0.044, respectively. The densities are normalized to the critical density using a Hubble constant of $H_0$ = 70~km~s$^{-1}$. 
  
 %================================================================ 
  \section{THE NATURE OF THE EXCESS BACKGROUND AT 1 TO 5 \mum}
  %================================================================

\subsection{Observations}

Figure 1 depicts the current limits and detections of the EBL in the 0.1 to 10 \mum\ wavelength region. These limits and detections were obtained by a variety of analyses and observational methods  including: (1) ground- and space-based measurements of integrated galaxy light (Madau \& Pozzetti 2000, open triangles; Bernstein, Freedman, \& Madore 2002, open diamonds); (2) direct measurements based on data obtained by the Diffuse Infrared Background Experiment (DIRBE) on board the {\it Cosmic Background Explorer} ({\it COBE}) satellite (Hauser et al. 1998; Dwek \& Arendt 1998; Arendt \& Dwek 2003, filled diamonds; Cambr\'esy et al. 2001, filled triangles; Wright \& Reese 2000; Gorjian Wright, \& Chary 2000; and Wright 2001, filled squares), and the Near Infrared Spectrometer (NIRS) on board the {\it Infrared Telescope in Space} ({\it IRTS}; Matsumoto et al. 2000, 2005, filled circles); and (3) extrapolated galaxy number counts (Totani et al. 2001, open circles). 
 The curves labeled "1", "2", and "3" in the figure are smooth polynomial fits representing three possible EBL spectra formed by star forming galaxies: a maximal one, designated EBL1, fitted to the Bernstein et al. (2002) data; a minimal one, designated EBL3, fitted to the lower limits defined by the galaxy number counts of Madau \& Pozzetti (2000); and an intermediate one, designated EBL2, fitted to the Totani et al. (2001) points. These polynomial fits include mid- and far-IR limits and measurements by Infrared Array Camera (IRAC) intsruments on board the {\it Spitzer} satellite (Fazio et al. 2004, open squares); Metcalfe et al. (2003, 15 \mum), Papovich et al. (2004, 24 \mum), Lagache et al. (2000, 100 \mum), and Hauser et al. (1998, 140, 240 \mum). Differences between the three backgrounds were found to be negligible beyond $\sim$ 10~\mum. 
 
The figure suggests that at $\sim$ 1 to 4~\mum\ the EBL intensity is higher than the integrated light from resolved galaxies (Madau \& Pozzetti 2000), or even their extrapolated intensity based on the evolutionary models (Totani et al. 2001). We will use the Totani et al. extrapolations as the baseline for the EBL intensity from normal galaxies, and designate hereafter this EBL2 component as GEBL. Furthermore, as suggested by  the figure, it seems hard to smoothly join the $\sim$ 1 to 4~\mum\  region of the EBL spectrum to that at shorter wavelengths while maintaining the general spectral shape of the integrated galactic starlight. Put differently, the GEBL formed by the unabsorbed stellar continuum emission, when normalized to fit the near-IR spectrum will require an integrated 0.1--10~\mum\ EBL intensity of about 100 \nwats\ (Hauser \& Dwek 2001). This value is considerably larger than  the total radiative energy released from star forming galaxies over the $z$ = 0 to 5 redshift interval, which is only about 50~\nwats after the fraction of starlight that has been absorbed and reradiated by dust has been subtracted. The near-infrared background light (NIRBL), defined hereafter as the $\sim$ 1 - 4 \mum\ diffuse light from which the extragalactic component (GEBL) has been subtracted, is therefore a distinct spectral component with an integrated intensity $\gtrsim$ 30~\nwats.

%------ figure 1 
  \begin{figure}
    \epsscale{0.5}
 \plotone{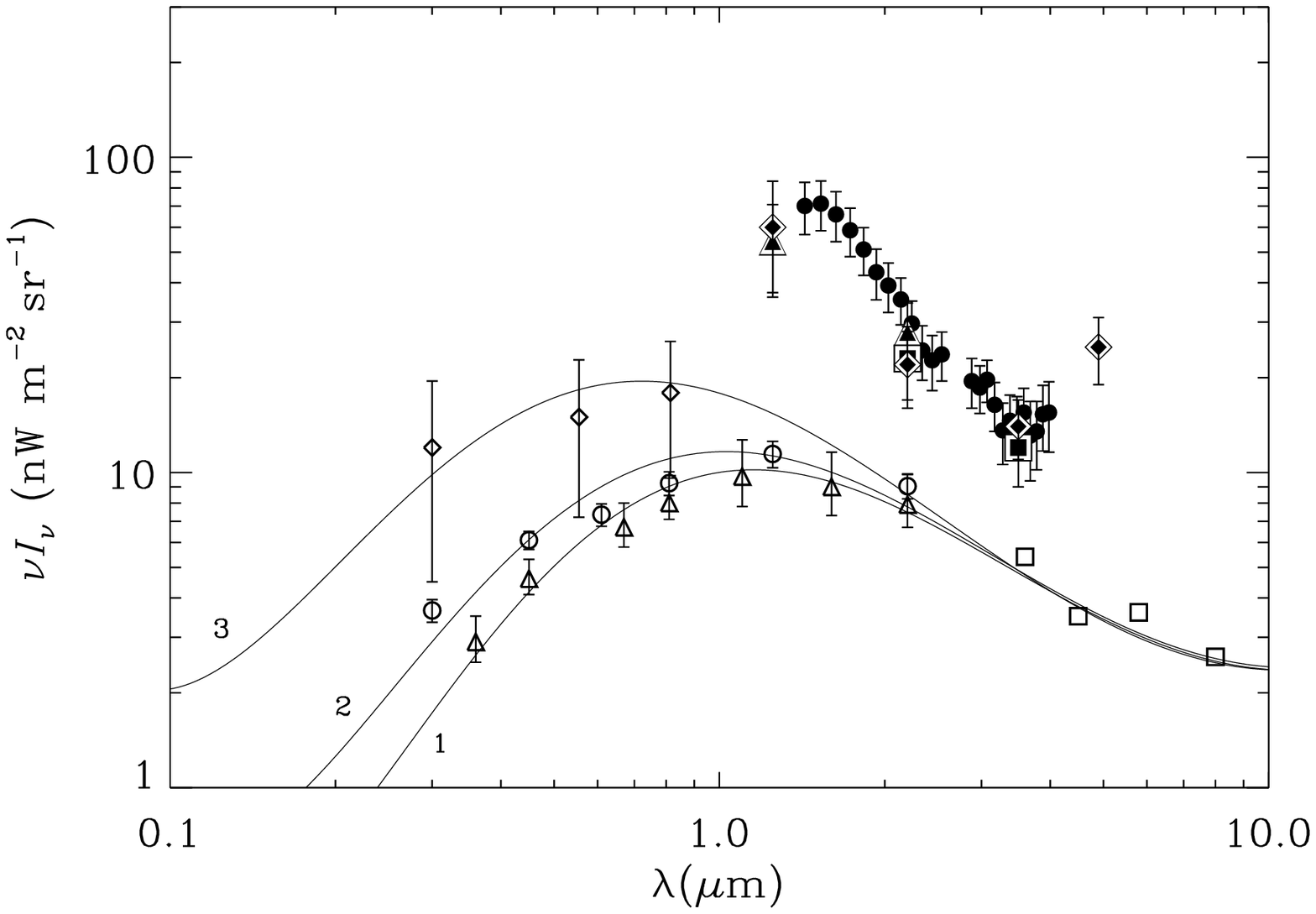}
 \caption{Limits and detections of the EBL. The three solid lines represent possible fits to the EBL spectrum generated by normal star forming galaxies. The curve labeled "2" (EBL2) is the nominal EBL spectrum attributed to these galaxies in this paper. References to the observations are in the text.}
   \epsscale{1.0}
 \end{figure}

\subsection{The Case for a Local Origin for the Emission}

We start with some cautionary notes concerning the attribution
of the excess DIRBE and IRTS emission to extragalactic sources.
First, there have been only two space missions {\it COBE}/DIRBE 
(Hauser et al. 1998) and {\it IRTS}/NIRS (Matsumoto et al. 2000, Matsumoto et al. 2005)
which have performed absolute measurements of the near-IR 
sky brightness over significantly large areas of the sky. The NIRS
measurements are much more limited in time and area than 
the DIRBE observations, but the total measured sky brightness does 
match DIRBE. While it is very encouraging that these two experiments 
are in excellent agreement, it will still be of great interest to have future 
experiments confirm these results with different instrumentation
and under different circumstances.

Second, the extragalactic interpretation of the origin  of the NIRBL hinges on the large discontinuity between the DIRBE and NIRS infrared measurements and the {\it Hubble} and ground-based UV-optical determinations of the EBL by Bernstein et al. (2002). Systematic errors in acquisition and analyses of these data sets may produce just such a discontinuity. Clearly, further observations covering the $\sim$ 0.1 to 10 \mum\ wavelength region with good spectral resolution would be valuable in revealing the true nature of the NIRBL.

Third, extraction of the NIRBL depends on accurate removal of
stellar and zodiacal foregrounds. 
The NIRBL appears to be only 
$\sim$10\% of the total near-IR sky brightness, so relatively small 
errors in the foregrounds may be significant. Hauser et al. (1998)
did not claim detections of the NIRBL because the uncertainties in the 
foreground models were too large. Subsequent analyses (e.g. Arendt \& Dwek 2003, and references therein) have aimed at reducing the uncertainties in the subtraction of foreground emissions.

In the near-IR there are two dominant foregrounds: \\
(1) {\bf Galactic stars} --
Their emission can be empirically removed by constructing spatial templates of the stellar emission (Dwek \& Arendt 1998), or by integrating NIR star counts (from high resolution observations)
and subtracting the flux from the low resolution DIRBE measurements at specific 
locations (Gorjian, Wright, \& Chary 2000; Wright 2001; Cambr\'esy et al. 2001). Both these methods are more robust than a statistical approach to the removal of the Galactic stellar foreground (Arendt et al. 1998).  
However, all these methods are susceptible to error if there exists a 
Galactic population of faint stars that have not been directly observed 
or used to constrain the statistical models. \\
 (2) {\bf Interplanetary dust particles} -- The strongest foreground to be removed is the zodiacal light produced by the scattering of sunlight by and thermal emission of interplanetary dust (IPD) particles. The zodiacal light is intrinsically diffuse, so unlike the stellar foreground it is not possible to resolve and subtract the individual sources regardless of spatial resolution and sensitivity. Most analyses of DIRBE and NIRS data have used the Kelsall et al. (1998) zodiacal light model or a closely related derivative (Hauser et al. 1998; Wright 1998; Wright \& Reese 2000; Matsumoto et al. 2005). This zodiacal light model was fit to the observed temporal variation of the observations. Thus the model is insensitive to any IPD component which is effectively isotropic and produces no significant variation in brightness over the range of elongations observed by DIRBE. Wright's versions of the zodiacal light model address this problem by placing some constraints in the mid-IR, but it's possible that there may be systematic errors in the model which affect the scattered light in the near-IR without affecting the thermal emission in the mid-IR. Components that are on a large enough scale to have small annual variations, or spherically distributed around the earth such that they have no annual variation, may be overlooked. 
 
There are several known spatial and temporal discrepancies between the Kelsall et al. (1998) model and the data. One, even noticeable at high ecliptic latitudes, is caused by the imperfect fitting of the earth-resonant ring and structures. 
Kelsall et al. (their Fig. 11a) show that at 12 \mum\ these errors can be up to 4\% of the model intensity,
which can lead to a 40\% error in the CIB estimate at this wavelength.
Another, appears as a periodic residual with a period matching the solar rotation period. These residuals are found to be strongly correlated with daily measures of solar activity, such as the Mg II index and Sunspot number (Kelsall et al. 1998; Kelsall, private communication). It is unknown whether these variations occur in the zodiacal cloud or in the earth's exosphere. 
These variations represent only $\sim 2\%$ of the IPD intensity at 1.25 and 3.5 \mum, but
without an understanding of the physical mechanism giving rise to the variation, it is hard to determine 
this component's contribution to a potential isotropic background.
 
Another problem arises from the interpolation of a model derived from the broad-band DIRBE observations to the narrow band NIRS observations. Matsumoto et al. (2005) needed adjustment factors as large as 25\% to get the Kelsall model to fit at some wavelengths. This suggests potential errors in the model that are similar to the levels of the residual emission. 
 
The greatest cause for concern 
is created by the fact that if we subtract the resolved extragalactic emission (EBL2) from the isotropic residual found by Matsumoto et al. (2005), then a simple scaling of their own zodiacal component produces an excellent fit to the residual emission.
Figure 2 shows the spectrum of the total sky intensity, the integrated starlight, and
the residual isotropic component, all normalized to the zodiacal light spectrum. While neither 
the total sky brightness not the starlight are exactly like the zodiacal light spectrum, 
the figure shows that the isotropic component found by Matsumoto et al. (2005) has a spectrum 
that is well matched by zodiacal light ($\chi^2 = 0.21$). Thus, the 0.232 mean ratio of
the isotropic / zodiacal intensity may imply a missing component of the zodiacal light model
which, if truly isotropic, would increase the zodical light intensity by $\sim25\%$ at ecliptic latitudes
$\beta \approx 50$ and smaller amounts at lower latitudes. 

The existence of such large missing isotropic component of the zodiacal  light is quite significant, but should not be totally surprising.  Kelsall et al. (1998) showed that at the
north Galactic pole ($\beta = 30$) the use of three alternate kernels for the geometry of 
the main IPD cloud can change the 2.2 \mum\ zodiacal light intensity by up to 14\%.
There is no certainty that there no other models with different geometries, or
additional components that may provide good fits to the temporal variations observed
by DIRBE while producing even larger mean brightnesses (and lowering
the level of any residual emission). This sort of systematic uncertainty is the most important 
for CIB studies, and it is also the hardest to estimate because of the difficulty in measuring
the absolute brightness of the zodiacal light alone. 

One observation that could eliminate these uncertainties in the absolute intensity of the zodiacal
light is measurement of scattered solar absorption lines at near-IR wavelengths.
Comparison of the relative depths of the lines in the solar and in the zodiacal spectrum can 
reveal the absolute intensity of the zodiacal light. Such measurements have been made in the
optical by the Wisconsin H-Alpha Mapper (WHAM) which has measured the profile of the 
scattered solar Mg~I $\lambda$5184 absorption line in the zodiacal light 
(Reynolds, Madsen, \& Moseley 2004). 

In summary, the NIRBL at 1.2 $< \lambda$(\mum) $<$~4 can be very well fit by a zodiacal dust spectrum, which is actually a nominally 
better fit than those of the Pop~III models presented later in this paper.  However, current observations suggest that the NIRBL spectrum drops significantly at wavelengths below $\sim$ 1~\mum. It is very unlikely that the zodiacal emission will contain such a discontinuity in the reflected solar spectrum. Consequently, if the apparent drop in the NIRBL spectrum is real, then an alternative non-local source is needed to account for its origin. 
In the following, we will therefore assume that the NIRBL is extragalactic, and examine the implication of its spectral shape and intensity for the formation of the first luminous objects in the universe. 

 %------ figure 2 
  \begin{figure}
    \epsscale{0.5}
 \plotone{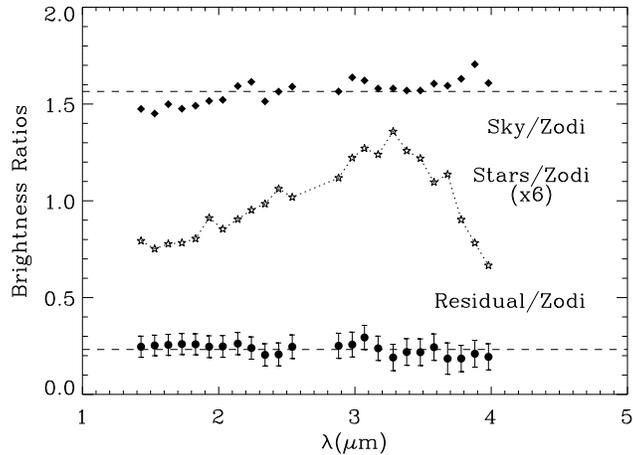}
 \caption{The spectrum of the total sky intensity, the integrated Galactic starlight, and
the residual isotropic component, all normalized to the zodiacal light (ZL) spectrum. Starting from the observed sky spectrum, which differs from that of the ZL, the subtraction of the contribution from Galactic stars, which is dramatically different from the ZL spectrum, leads to a residual spectrum which is very well matched by that of the ZL. The residual spectrum represents 23.2\% of the total ZL intensity. }
   \epsscale{1.0}
 \end{figure}

%================================================================ 
  \section{A COSMOLOGICAL ORIGIN OF THE EXCESS EMISSION}
  %================================================================
 
\subsection{Global Energetics and Metallicity Considerations}
Before pursuing more detailed modeling, it is instructive to estimate the cosmic energy release required to account for the total NIRBL intensity. We assume that the comoving luminosity density is given by:
\begin{equation}
j(z) = j(z_f) \left({1+z\over 1+z_f}\right)^{-2}
\end{equation}
where $z_f$ is the redshift corresponding to the last epoch of this energy release.
The total EBL intensity is given by:
\begin{equation}
I_{EBL} = \left({c\over 4 \pi}\right)\ \int_{z_f}^{\infty} j(z)\ \left|{dt\over dz}\right|\ {dz\over 1+z}
\end{equation}
where
\begin{eqnarray}
\left|{dt\over dz}\right| & = & H_0^{-1} \left[(1+z) E(z)\right]^{-1} \\ \nonumber
E(z) & \equiv & \left[ \Omega_{\Lambda} + (1+z)^3 \Omega_m \right]^{\frac{1}{2}}
\end{eqnarray}

Figure 3 depicts the value of $j(z_f)$, needed to produce a given background intensity for different values of $z_f$. 
The integrated NIRBL intensity is  30 \nwats in the $\sim$ 1.4--4 \mum\ NIRS wavelength region. The figure shows that for $6 < z_f < 9$, a luminosity  density of $j =  (0.7 - 1.8) \times 10^{11}$ \Lsun\ Mpc$^{-3}$ is needed to reproduce this intensity. Smaller values of $z_f$ need lower luminosity densities.
We note that in this exercise, only the 1.4--4~\mum\ NIRBL intensity is being considered. Later, we present more detailed models of Pop III emission which produce emission outside this wavelength range, and thus lead to somewhat larger intensities and luminosity densities.

Released over an effective period of $\Delta t$, the corresponding comoving energy density is $\epsilon$ is:
\begin{equation}
\epsilon({\rm erg\ Mpc}^{-3}) \equiv j \times \Delta t 
\end{equation}
where,
\begin{equation}
\Delta t  \equiv  \int_{z_f}^{\infty} \left({j(z) \over j(z_f)}\right)\ \left|{dt\over dz}\right|\ dz
\end{equation}
Values of $\Delta t$ for  $z_f$ = \{6,\ 7,\ 9\} are: $\Delta t$(Myr) = \{413,\ 339,\ 242\}, giving 
\begin{eqnarray}
\epsilon ({\rm erg\ Mpc}^{-3}) & = & 3.4 \times 10^{ 60} \qquad {\rm for} \ z_f = 6 \\ \nonumber
& = & 3.6 \times 10^{ 60} \qquad {\rm for} \ z_f = 7 \\ \nonumber
& = & 5.2 \times 10^{ 60} \qquad {\rm for} \ z_f = 9
\end{eqnarray}

\noindent
The release of this energy density by hydrogen burning in stars will convert a comoving mass density 
\begin{equation}
\rho_M \equiv {\epsilon \over \eta c^2} \approx 3.2\times 10^8\ M_{\odot}\ {\rm Mpc}^{-3}
\end{equation}
of hydrogen into helium and heavier elements, where $\eta$ = 0.007 is the energy conversion efficiency for nuclear energy generating reactions.  Injected into the intergalactic medium at redshifts $\gtrsim$ 7, this will create an early enrichment of 
\begin{equation}
-\Delta X = \Delta Y + \Delta Z = {\rho_M\over \rho_b} \approx 0.052
\end{equation}
 where $X$, $Y$, and $Z$ are, respectively, the cosmic mass fractions of hydrogen, helium, and heavier elements, and $\rho_b = 6.0\times 10^9$~\Msun\  Mpc$^{-3}$ is the comoving baryonic density. This value is comparable the current enrichment in helium ($\Delta Y \approx$ 0.04) and heavy elements ($\Delta Z \approx$ 0.02) since the Big Bang, leaving no room for significant production of metals by Pop~II and Pop~I stars. 
 
To avoid this early overproduction of He and metals, it has been suggested by various authors (e.g. SBK) that Pop~III stars should be very massive objects with masses above 260~\Msun, which collapse directly into a black hole at the end of their lifetime (Heger et al. 2003). These suggestions are also supported by models showing that Population~III stars could have been more massive than present day stars because of the inefficient fragmentation of the primordial condensations (Kashlinsky \& Rees 1983, Bromm \& Larson 2004). We will therefore assume that the Pop~III stars giving rise to the NIRBL are sufficiently massive so that they directly form black holes at the end of their lifetimes. The added advantage of these massive stars is that they are fully convective and that over their lifetimes they convert all their initial hydrogen into helium and heavier elements. Therefore, they require the minimal amount of baryons to be incorporated into stars in order to generate the required NIRBL intensity. The baryonic mass fraction that needs to be converted into stars is then given by: 
 \begin{equation}
f_b = {\epsilon \over \eta \rho_b c^2} = {\rho_M\over \rho_b} \approx 0.052 = - \Delta X
\end{equation}
This corresponds to a comoving stellar mass density of 
\begin{equation}
\rho_* = \rho_M = 3.2\times 10^8\ M_{\odot}\ {\rm Mpc}^{-3}
\end{equation}
 Generated over an effective period of $\Delta t$ this gives a mean comoving star formation rate of $\sim$1 \Msun\ \yr\ \Mpc.
   
All rates and densities derived above are larger by factors of 2 to 4 than those derived from models in which Pop~III stars form at a rate proportional to the collapse rate of haloes in a cold dark matter universe with a star formation efficiency of about 40\% (Bromm \& Loeb 2002, Madau \& Silk 2005). However, the discrepancies may be even larger because: (1)  the star formation rate derived above represents a {\it lower limit} on the formation rate of Pop~III stars that is required to generate the NIRBL intensity, since it assumes that all the mass incorporated into stars is transmuted by nuclear reactions into heavier elements; and (2) the NIRBL intensity included only the radiation detected in the 1.4 to 4 \mum\ band, and any emission from Pop~III stars outside this band will increase the required energy production and star formation rates. 

In the following we present more detailed models of the contribution of Pop~III stars to the EBL.   

 %------ figure 3
  \begin{figure}
  \epsscale{0.5}
 \plotone{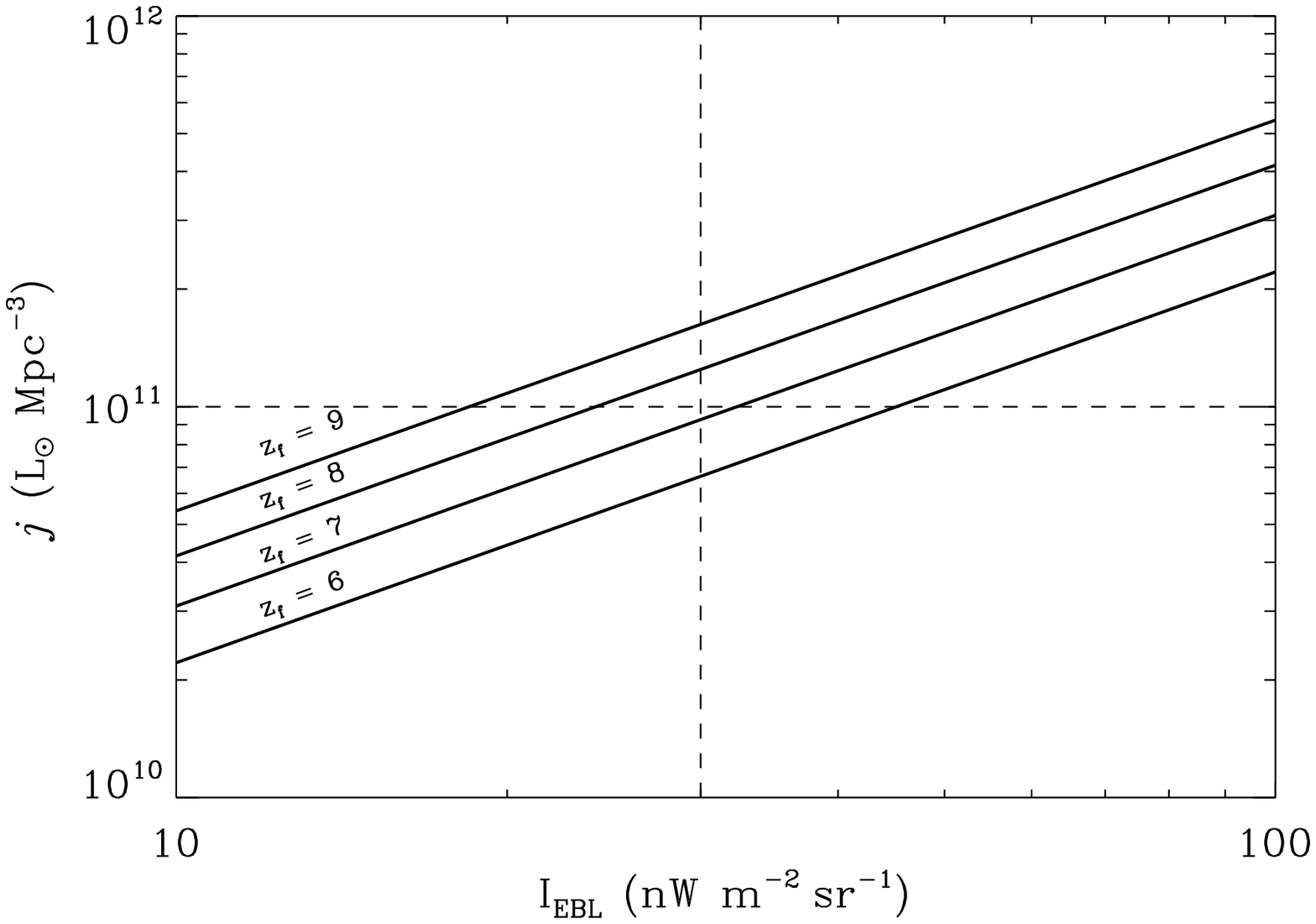}
 \caption{The comoving luminosity density at redshift, $j(z_f)$, needed to produce a given EBL intensity, for different values of $z_f$, the final epoch of energy injection. The luminosity density was assumed to have a $(1+z)^{-2}$ dependence at $z > z_f$. Calculations for different values of $z_f$ are represented by solid lines. The vertical dashed line corresponds to the value of the NIRBL intensity integrated over the 1.4 - 4~\mum\ interval.}
   \epsscale{1.0}
 \end{figure}

 \subsection{Models of Emission from Pop~III Stars}

\subsubsection{Stellar Parameters}

Calculations are considerably simplified when all stars have masses, $M$, in excess of $\sim$ 100~M$_{\odot}$, beyond which stars are dominated by radiation pressure. Their luminosity is approximately given by the Eddington limit, $L_{{\rm Edd}}$/L$_{\odot}$ = 3.3$\times10^4~M$/\Msun; their effective temperature is roughly independent of mass and approximately equal to 10$^5$~K (Bond, Arnett, \& Carr 1984); their spectrum is characterized by that of a blackbody at that temperature (Bromm, Kudritzki, \& Loeb 2001); and their lifetime, given by $\tau_* = \epsilon Mc^2/L \approx 3.2\times 10^6$~yr, is independent of the stellar mass. Furthermore, if they have masses in excess of 260~\Msun, they end their lives as collapsed objects, preventing the overproduction of metals at high redshifts. For the purpose of our study, we will adopt the same stellar parameters as those used in the study of SBK: a stellar mass of $M_*$ = 10$^3$~M$_{\odot}$, yielding a luminosity $L_*$ = 3.3$\times10^7$~\Lsun, and a stellar spectrum characterized by that of a 10$^5$~K blackbody  

\subsubsection{The Spectrum of an Ionization Bounded Primordial H~II Region}

The ionizing photons emanating from the stars can be either absorbed in the remnant nebula from which the stars formed or escape into the intergalactic medium (IGM) if the nebula is not massive enough. As shown by SBK, the two cases will leave different imprints on the NIRBL. A larger  EBL intensity is generated when all the ionizing photons are assumed to be locally absorbed compared to the case in which the ionizing radiation escapes into the IGM.  In the following we assume that all the ionizing photons are locally absorbed, creating an ionization bounded H~II region around the star. Given a NIRBL intensity, this assumption will lead to a lower limit on the Pop~III star formation rate.

We used CLOUDY (Ferland 1996) to calculate the spectral signature of a Pop~III star and its surrounding H~II region assuming that all the stellar ionizing photons are absorbed in the surrounding nebula. The nebula was assumed to have primordial composition (H and He mass fractions of 0.76 and 0.24, respectively), and a hydrogen number density of 10$^4$~cm$^{-3}$. The radiation emerging from the nebula comprises three spectral components: (1) a stellar continuum component consisting of photons that were not absorbed by the ambient gas; (2) a nebular continuum emission component; and (3) a nebular line emission component. The emission lines emerging from the H~II regions are scattered by the ambient IGM, and we have used SBK's approximation (their eq. 15) to calculate the line profile originally derived from Monte-Carlo simulations by Loeb \& Rybicki (1999). Figure 4 depicts the three emission components in the rest-frame of the star, assumed (for the purpose of calculating the scattered line profile) to have formed at $z$ = 15. 

The mass of the ionized nebula is about 1.4$\times10^4$~\Msun, which is the minimum mass required to create an ionization bounded H~II region for the given stellar parameters. This will imply a star formation efficiency $\eta_* \approx 1/(1+1.4) \approx$ 0.41. The H~II region is therefore ionization bounded for $\eta_* \lesssim 0.41$. 
Table 1 presents the parameters of the H~II region relevant to our calculations.

%====== figure 4 ====== 
  \begin{figure}
    \epsscale{0.5}
 \plotone{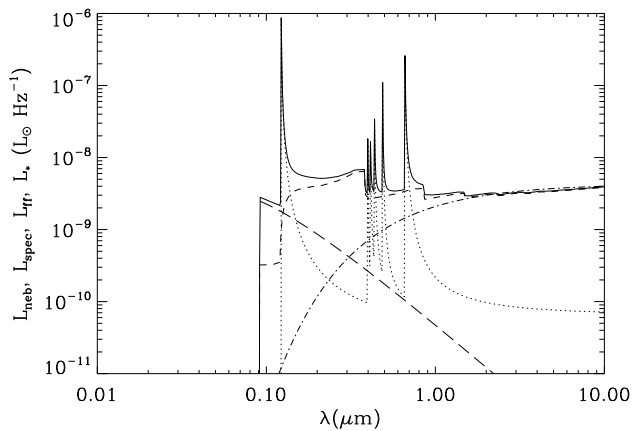}
 \caption{Components of the Pop~III source spectrum: (1) escaping non-ionizing stellar radiation ($L_{\nu}^*$, long dashes); (2) nebular continuum emission ($L_{\nu}^{cont}$, dashed line); and (3) nebular line emission ($L_{\nu}^{lines}$; dotted line), represented here by the Ly$\alpha$ and Balmer series recombination lines. Also shown in the figure is the free-free contribution (dashed-dotted line) to the nebular continuum emission component. The source luminosity was calculated for a 10$^3$~\Msun\ star radiating at the Eddington luminosity with a blackbody spectrum at $T$ = 10$^5$~K. Line spectra were broadened to account for scattering in the IGM, assuming an injection epoch of $z$ = 15. Total luminosities in each component are presented in Table 1.}
   \epsscale{1.0}
 \end{figure}

\subsubsection{Global Parameters}

At electron densities of 10$^4$~cm$^{-3}$ the recombination timescales are short compared to the stellar lifetime of $\tau_* \approx 2\times 10^6$~yr. The creation of the NIRBL requires therefore the continuous formation of stellar objects over a yet to be determined redshift interval \{$z_{min}$, $z_{max}$\}. 
The contribution of the star and H~II region then creates a spectrum with an intensity, $\nu I_{\nu}(\nu)$, given by:
\begin{equation}
\nu I_{\nu}(\nu)  =   \left({c \over 4 \pi}\right) \int_{z_{min}}^{z_{max}} n_*(z) \left[L_{\nu}^*(\nu',z) + L_{\nu}^{cont}(\nu',z) +
L_{\nu}^{lines}(\nu',z)\right]\ \left|{dt\over dz}\right|dz
\end{equation}
 where the terms in the square brackets represent, respectively, the spectral luminosity density (\Lsun~Hz$^{-1}$) of the escaping 
 stellar, nebular continuum, and line emission components, \nun$' \equiv$ \nun(1+$z$) is the frequency of the emitted photon and \nun\ its observed one, and $n_*(z)$ is the comoving number density of Pop~III stars, assumed to have a $z$ dependence of the form:
\begin{equation}
n_*(z) = n_0\left({1+z\over 1+z_{min}}\right)^{\alpha}
\end{equation}
The limiting redshifts $z_{min}$, $z_{max}$,  and $n_0$, and $\alpha$ are adjustable parameters chosen to give the best fit of the model spectrum to the observed NIRBL.
 
To illustrate the sensitivity of the NIRBL spectrum to the model input parameters we calculated $\nu I_{\nu}(\nu)$ for several values of $z_{min}$, $z_{max}$. All models were calculated for a value of $\alpha$ = -2.0, which provided a good fit to the slope of the NIRBL spectrum. For each choice of $z_{min}$ and $z_{max}$, $n_0$ is determined by a least-squares fit to the NIRBL. 
Calculations were performed for three values of $z_{min}$: the lowest value of  $z_{min}$ = 6, was chosen so that the short-wavelength edge of the redshifted \lya\ line will be just above the EBL determination of Bernstein et al. (2002). This choice of $z_{min}$ has the added attractive feature of redshifting the \ha\ line into the 4.9 \mum\ DIRBE bandpass. A maximum value of $z_{min}$ = 9 was chosen so that the edge of the \lya\ line will coincide with the 1.25~\mum\ EBL intensity, and finally, calculations were also performed for an intermediate value of $z_{min}$ = 7. Salvaterra \& Ferrara (2003a) obtained "hard" limit of $z_{min}$ = 8.8,  motivated by the apparent drop in NIRBL between the NIRS data and the DIRBE J(1.25 \mum) band intensity. However, the drop in the EBL intensity at this wavelength is statistically insignificant, and should not be used to constrain the value of $z_{min}$. The value of $z_{max}$ was varied from 15 to a maximum value of 30, since model results did not vary significantly for larger redshifts. 

\subsection{Model Results}
Figure 5 summarizes the results of our calculations. All figures display the three EBL spectra that can be attributed to star formation at $z \lesssim$~5. Also shown in the figure are the spectral contribution of the non-ionizing stellar (long dashes), the nebular continuum (short dashes), and the nebular line (dotted line) emission components to the NIRBL intensity. The thick solid line represents the sum of these emission components added to the EBL2 intensity. In addition to the values of $z_{min}$ and $z_{max}$, the figures also give the value of $n_0$ in \Mpc, and the reduced $\chi^2$ of the fit to the NIRS data points.

%====== figure 5 ====== 
  \begin{figure}
 \plottwo{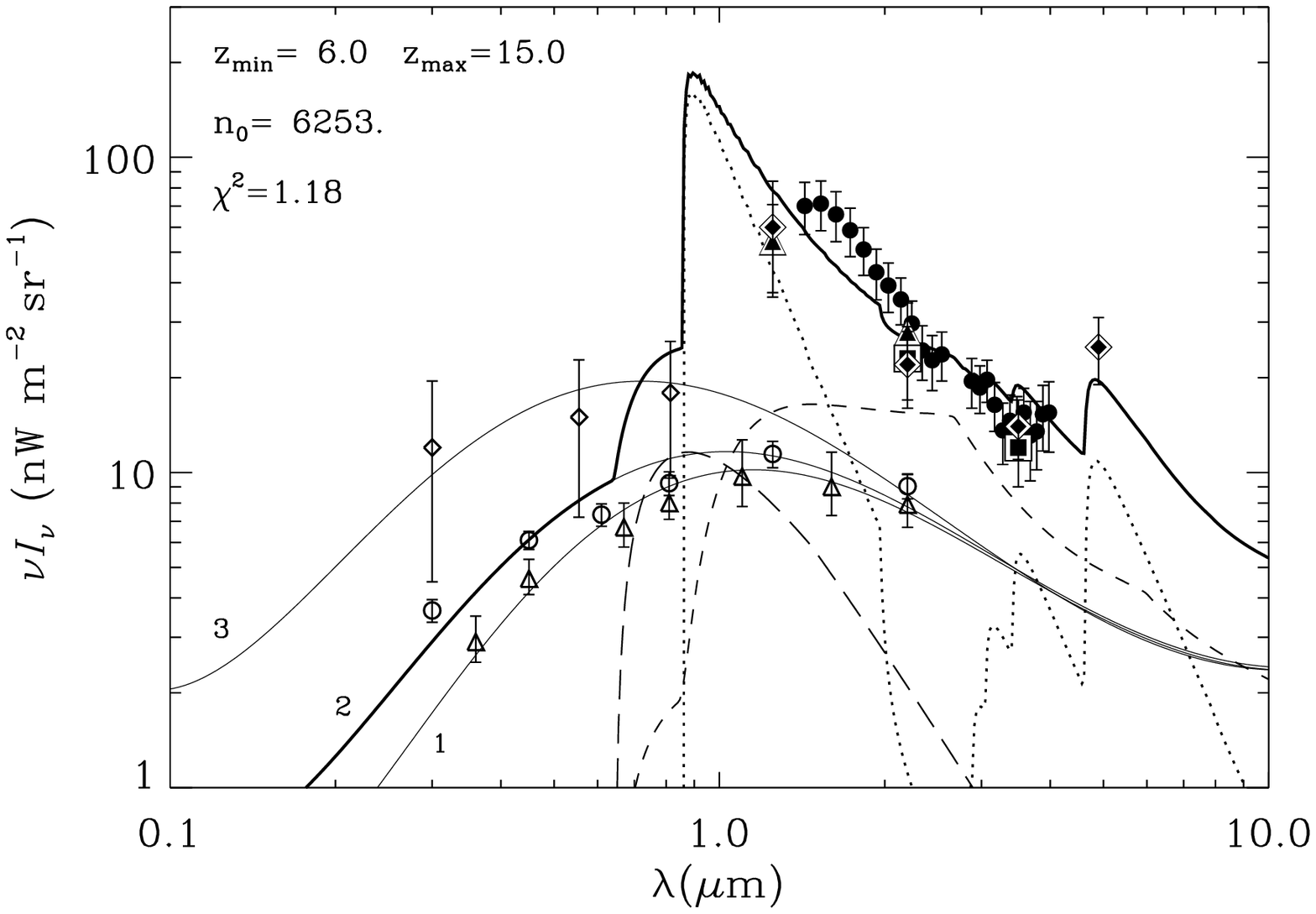}{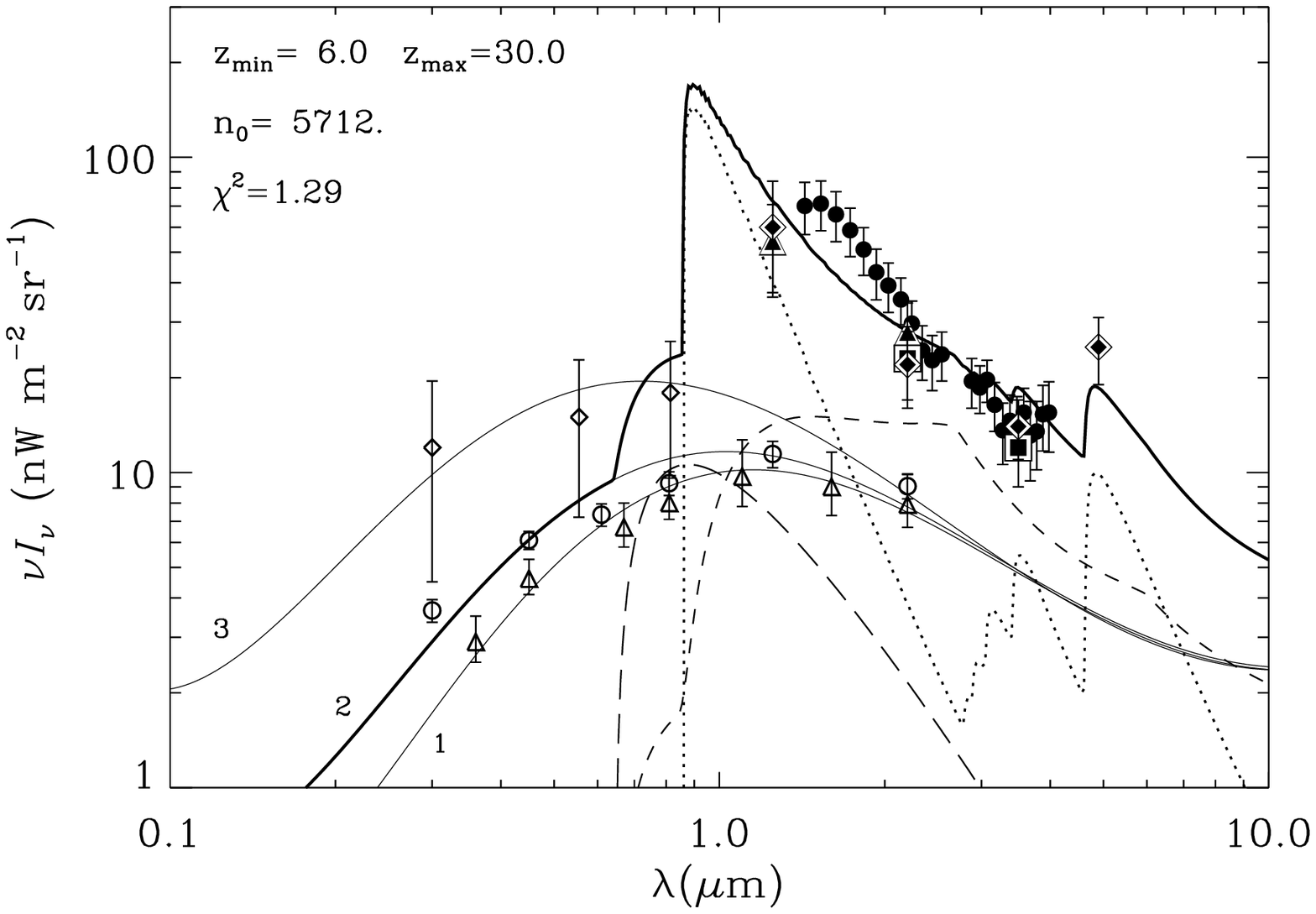}\\
 \plottwo{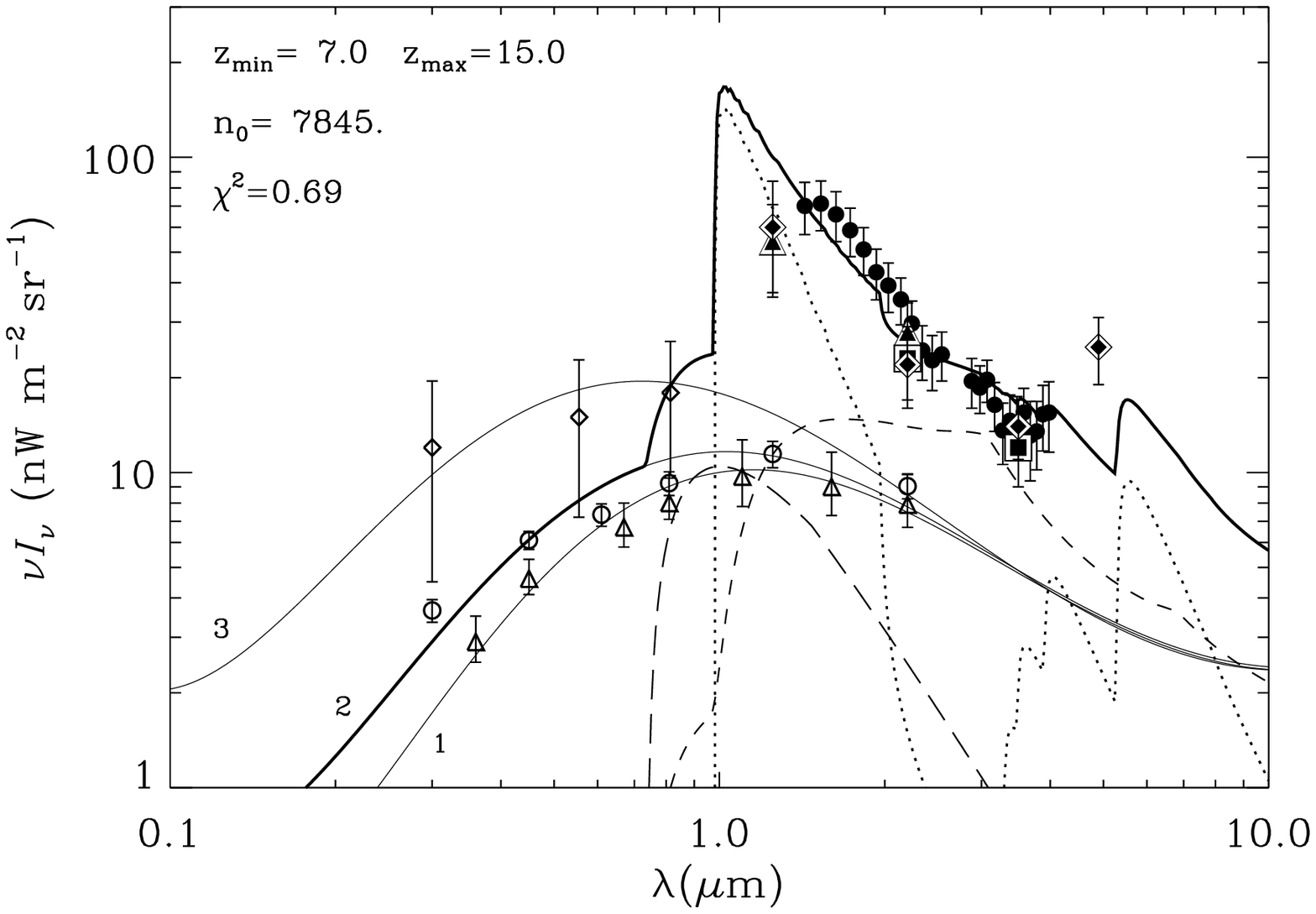}{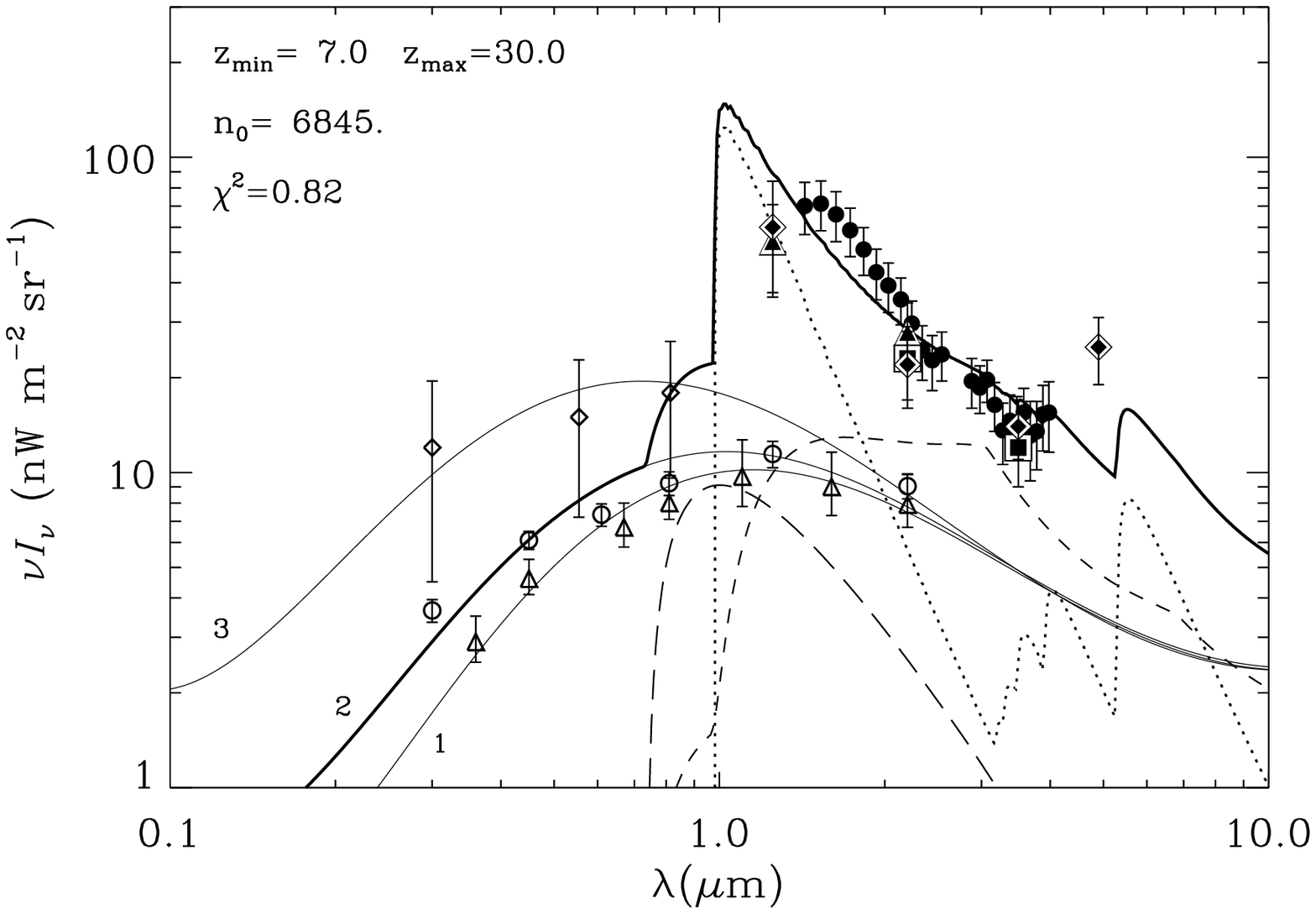}\\
 \plottwo{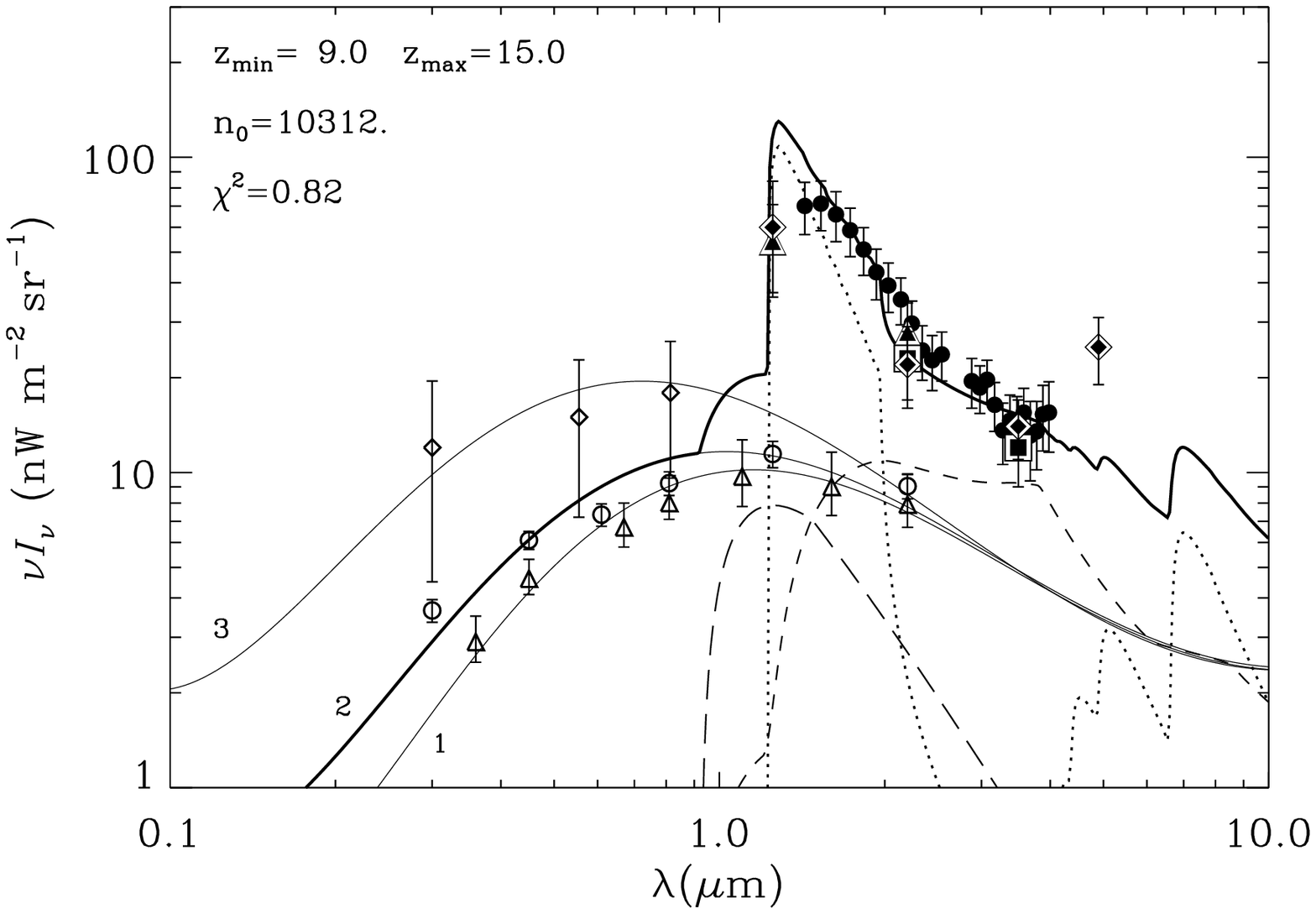}{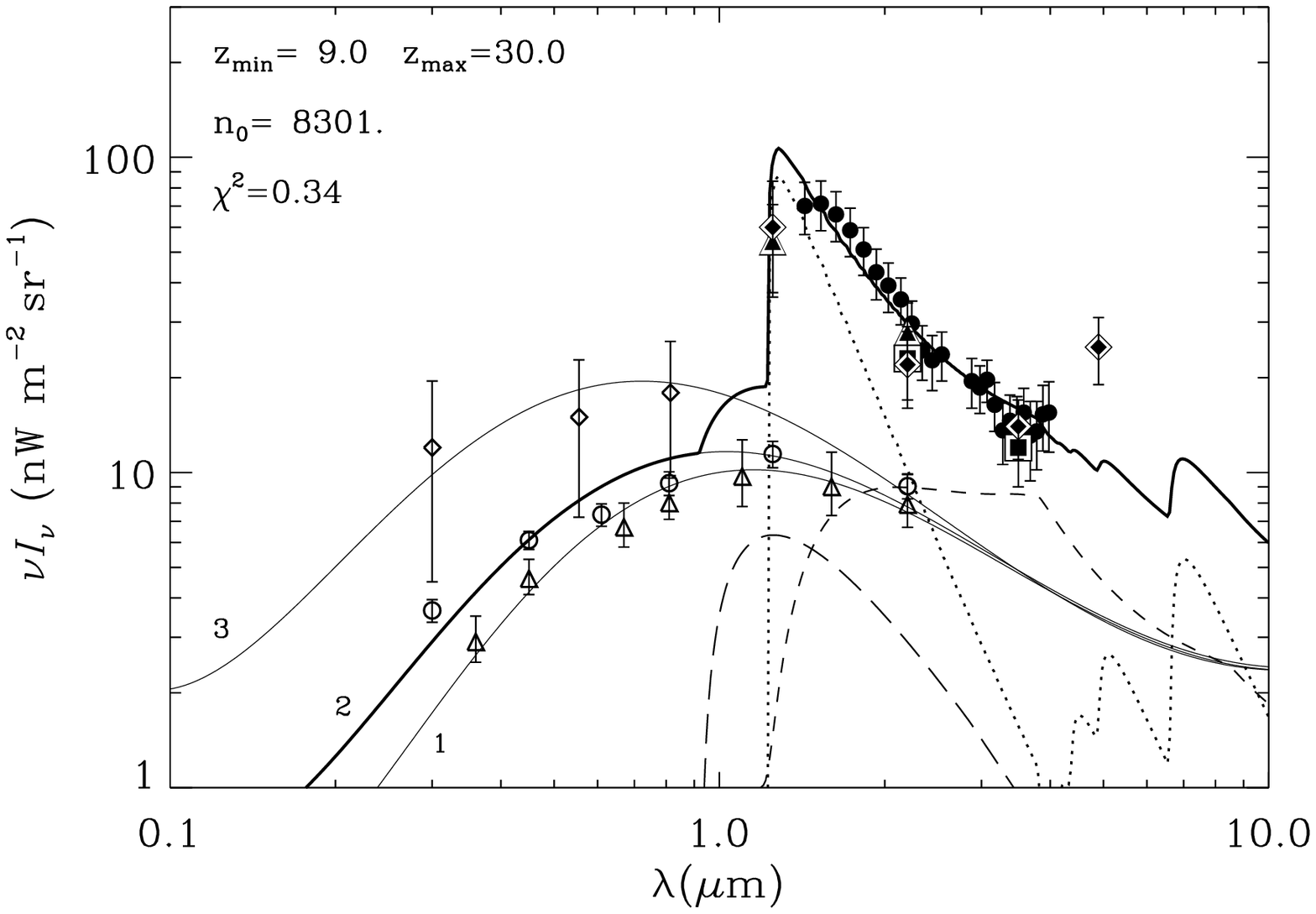}\\
 \caption{The imprint of Pop III stars on the near-IR EBL for different values of \{$z_{min}$, $z_{max}$\}. Curves labeled 1, 2, and 3 are different fits to the EBL formed by normal star forming galaxies. The heavy solid line represents the sum of the Pop~III contribution and EBL2.  Also shown in the figure are the three Pop~III emission components: the non-ionizing photons escaping from the nebula (long dashed line); the nebular continuum and free-free emission (dashed line); and the nebular line emission (dotted line). The thick solid line is the sum of all three components added to EBL2. All models were calculated for a star formation rate $n_*(z) = n_0[(1+z)/(1+z_{min})]^{\alpha}$, with $\alpha =-2.0$. The value of $n_0$ is given in units of \Mpc.}
 \end{figure}

Figures 5a,b (top two panels) shows the results for $z_{min}$ = 6. In spite of the attractive result that the model can account for the upturn of the DIRBE 4.9~\mum\ intensity, it otherwise provides a poor fit to the NIRBL at shorter wavelengths. Figures 5e,f (bottom two panels) depict the resulting Pop~III spectra for \zmin = 9, and different choices of \zmax. For this value of \zmin, none of the Balmer lines contributes significantly to the 4.9 \mum\ intensity. The H$\beta$ line falls in that wavelength band, however, the line intensity and the intensity of the nebular emission are significantly reduced compared to the \zmin\ = 6 case, because the epoch of their formation has been shifted to higher redshifts. Consequently, the Pop~III spectrum never contributes significantly to the 4.9 \mum\ emission. For \zmax\ = 15, the long wavelength edge of the \lya\ line occurs around 2~\mum, and the spectrum drops somewhat below the observed NIRBL. A value of \zmax\ = 30 provides a somewhat better fit to the data. This case illustrates the importance of the NIRS wavelength coverage in constraining the high-$z$ end of the redshift interval. With only the DIRBE data, both fits would seem equally acceptable. 
Figures 5c,d (two middle panels) show the model results for \zmin\ = 7, and different values of \zmax. The figures show an improvement in the fit as \zmax\ decreases from 30 to 15.
 Overall, the best fits to the NIRBL are provided by the following models, designated NIRBL7 and NIRBL9, characterized by the following set of parameters: 
 \begin{eqnarray}
 \{z_{min},\ z_{max},\ \alpha,\ n_0\} & = & \{7,\ 15,\ -2.0,\ 7.8\times 10^3\ {\rm Mpc}^{-3}\} \qquad {\rm model\ NIRBL7}\\ \nonumber
  & = & \{9,\ 30,\ -2.0,\ 8.3\times 10^3\ {\rm Mpc}^{-3}\} \qquad {\rm model\ NIRBL9}
\end{eqnarray}
Figure 6 depicts the spectral signature of the best fits of the different models to the NIRBL.

 %------ figure 6
  \begin{figure}
    \epsscale{0.5}
 \plotone{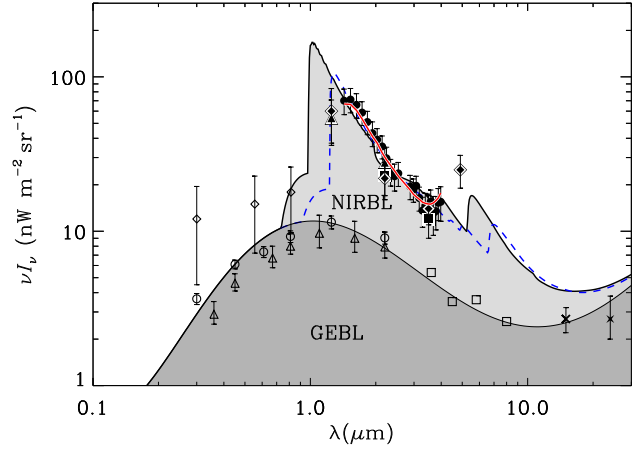}
 \caption{Comparison of current limits and detections of the extragalactic background light (EBL) to the spectrum generated by normal star forming galaxies (GEBL = EBL2 in Figure 1, dark shaded area), and the excess near--IR background light (NIRBL, lightly shaded area). The limits and detections are based on data and analyses including ground and space based measurements summarized by Hauser \& Dwek (2001) with additional data from Matsumoto et al. (2005),  Cambr\'esy et al. (2001), Fazio et al. (2004), and Papovich et al. (2004). The open circles represent extrapolated galaxy number counts (Totani et al. 2001). The solid blue line represents the sum of the Pop III model NIRBL7 and the GEBL, and the dashed blue line that of model NIRBL9 plus the GEBL. The red curve depicts the spectrum of the zodiacal light (Matsumoto et al. 2005), scaled by a factor of 0.23 to provide a fit to the NIRS data.}
   \epsscale{1.0}
 \end{figure}

 %================================================================ 
  \section{COSMOLOGICAL IMPLICATIONS}
  %================================================================

\subsection{The Formation Rate of Pop~III Stars}
Model fits to the NIRBL provide the comoving number density of Pop~III stars, $n_*(z)$ as a function of redshift.
The formation rate, $\dot \rho_*(z)$, of Pop~III stars for $z \ge z_{min}$ is given by:
\begin{eqnarray}
\dot \rho_*(z) & = & n_*(z) \times \left({M_*\over \tau_*}\right) \\ \nonumber
 & = & n_0  \times \left({M_*\over \tau_*}\right) \left({1+z\over 1+z_{min}}\right)^{-2}  \\ \nonumber
 & \equiv & \dot \rho_0\  \left({1+z\over 1+z_{min}}\right)^{-2} 
\end{eqnarray}
 Figure 7 compares the formation rate of Pop~III stars for the two different Pop~III formation scenarios to that of stars in normal galaxies, for the model output parameters given in Figure 5 (see also Table 2). 
 The figure shows that the formation rate of Pop~III stars is higher by a factor of ten than the star formation rate (SFR) in normal galaxies, with an abrupt drop around the reionization redshift \zmin. 

%====== figure 7 ====== 
  \begin{figure}
  \epsscale{0.5}
 \plotone{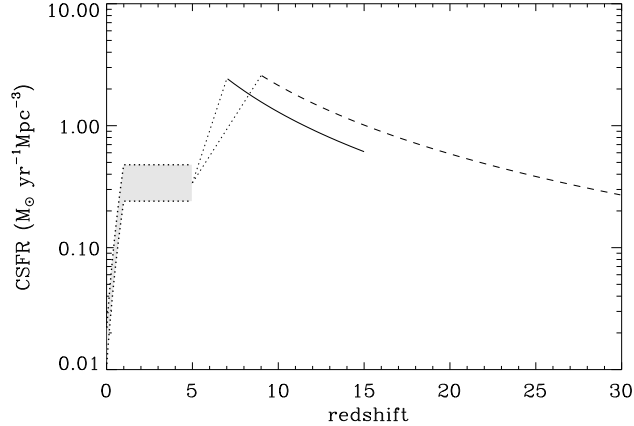}
 \caption{The star formation rate due to normal galaxies (shaded area; see Gabasch et al. 2004 for references), and Pop~III stars. The solid and dashed lines represents models NIRBL7 and NIRBL9, respectively, characterized by model parameters given in eq. (6), and spectra given in Figure 5.}
  \epsscale{1.0}
 \end{figure}
 
In the hierarchical model for structure formation in a cold dark matter (CDM) dominated universe, galaxies form out of the cooling gas streaming into the potential well of collapsed dark halos, which evolve with redshift through a series of hierarchical mergers. The star formation rate is then equal to the growth rate of the mass of collapsed haloes that are above some critical mass, $M_{crit}$,  times the product of the fraction of the total mass of dark matter that in baryonic form and the star formation efficiency $\eta_*$. The latter quantity depends on details of the state (ionized, atomic, or molecular) of the primordial gas, and for later generations of star, the metallicity of the gas. Bromm \& Loeb (2002) derived a peak star formation rates between $\sim$ 0.3 and 0.6~\Msun\ \yr\ \Mpc, depending on the phase (atomic or molecular) of the collapsing gas, at $z \approx$ 10. Our results show a similar functional behavior, but with a SFR larger by factors of 4 to 10, which is a direct consequence of attributing the NIRBL to the energy output from these stars. 

\subsection{The Fraction of Baryons Converted to Stars}
The total comoving mass density of baryons converted to stars is given by $\rho_* = \dot \rho_0\ \Delta t$, where $\Delta t$ is given by eq. (5).
The fraction of baryons in the universe that need to be   converted into Pop~III stars in order to account for the NIRBL intensity is given by $\rho_*/\rho_b$, where and is equal to 10--12\% (see Table 2). 

\subsection{The Energy Released per Baryon}
The total energy density emitted by the Pop~III stars in order to produce the NIRBL is given by 
\begin{equation}
\epsilon_* \equiv j_0\ \Delta t = n_0\ L_*\ \Delta t \ \ ,
\end{equation}

\noindent
which has an average value of $9\times 10^{60}$ erg \Mpc. Table 2 gives the value of $\epsilon_*$ for the two Pop~III formation scenarios. The value of $\tilde E_b$, the energy released by Pop~III stars per baryon in the universe, ranges from 670 to 820 keV/baryon.

\subsection{The Ionization of the Intergalactic Medium by Pop~III Stars}
Throughout this work we assumed that all the ionizing photons are absorbed in the surrounding nebula, with none escaping into the intergalactic medium. Such scenario is inconsistent with the observational evidence that most of the IGM was ionized at $z \gtrsim$~7. We will therefore relax this assumption, and calculate the fraction of the ionizing photons that need to escape the nebula in order to ionize the IGM at this redshift.

The number of ionizing Lyc photons, $N_{Lyc}$ emitted by a star with a blackbody spectrum of temperature $T$ and  a luminosity $L$ is given by:
\begin{eqnarray}
{dN_{Lyc}\over dt} & = & \left({15\over \pi^4}\right)\ \left({L\over kT}\right)\int_{x_0}^{\infty} {x^2dx\over \exp(x)-1} \\ \nonumber
 & = & 2.38\times 10^{51} \ {\rm s}^{-1}
\end{eqnarray}
 where $x \equiv h\nu/kT$, $x_0 = x$ at the Lyman limit, and the second line was calculated for  $L = 3.3\times10^7$~\Lsun, and $T = 10^5$~K.
 
To keep most of the IGM ionized, a fraction of these photons need to leak out of the nebula, ionizing the surrounding medium so that the Str\"omgren spheres around these stars will have a volume filling factor of about unity.
For a perfectly uniform distribution of Pop~III stars, the average volume occupied by each star is given by $n_0^{-1} \approx$ 1/8000~Mpc$^3$. The injection rate of ionizing photons required to keep this volume fully ionized is given by:
 \begin{eqnarray}
{dN_{ion}\over dt} = n_i n_e \alpha V
\end{eqnarray}
where $\alpha \approx 4\times10^{-13}$~cm$^3$~s$^{-1}$ is the recombination factor calculated for a gas temperature of $10^4$~K, and $n_i$, $n_e$ are, respectively, the number density of ions and electrons of the medium. The baryonic number density at $z$ = 7 is $n_b = 1.3\times 10^{-4}$~cm$^{-3}$. So for a fully ionized IGM we get that
 \begin{equation}
{dN_{ion}\over dt} = 2.5\times10^{49} \ {\rm s}^{-1} \qquad.
\end{equation}
Comparison to the production rate of ionizing photons given by eq. (9), we find that for a uniform distribution of Pop~III stars, only 1\% of the ionizing stellar photons need to escape the nebula in order to create a fully ionized IGM. This small fraction will have no noticeable effect on our calculations of the NIRBL, which assumed  that all the ionizing photons are locally absorbed. 
  
 %==================================
 \section{OTHER INSIGHTS INTO THE ORIGIN OF THE NIRBL}
 %==================================
 
\subsection{Spatial Fluctuations in the EBL}

Random spatial fluctuations measured in the background emission for
DIRBE, 2MASS and NIRS (Kashlinsky \& Odenwald 2000; Kashlinsky et al. 2002;
and Matsumoto et al. 2005) have been presented as evidence for the extragalactic nature of  the NIRBL (Magliocchetti, Salvaterra, \& Ferrara 2003a).
These fluctuations are found to be in excess of those expected from
instrumental noise and local IPD or galactic stars, and even
larger than the fluctuations expected from normal galaxy populations.
On smaller spatial scales ($<1\arcdeg$), the angular correlations
or angular power spectra of these fluctuation are in rough agreement
with models (Kaslinsky et al. 2004). However, as noted by Matsumoto et al. (2005) fluctuations on larger scales have amplitudes larger than those predicted by the present Pop~III models. It is therefore presently ambiguous whether the observed fluctuations provide evidence for or against the extragalactic nature of the NIRBL. 

\subsection{The effect of the NIRBL on the Absorption-Corrected TeV Spectrum of Blazars}
Additional claims for the extragalactic nature of the NIRBL are based on the EBL opacity to very high energy $\gamma$-rays. The $\gamma$-ray photons from blazars interact with EBL photons producing electron-positron pairs. The NIRBL, if extragalactic will therefore produce a discontinuity in the TeV opacity towards a blazar which, in principle, will leave its imprint on  its intrinsic spectrum. Mapelli, Salvaterra, \& Ferrara (2004) have suggested that the observed ÒSÓ--shaped spectrum of H1426+428 is the manifestation of this absorption feature, assuming its intrinsic spectrum rises as $E^2 dN/dE \propto E^{0.4}$. However, an intrinsic spectrum in which $E^2 dN/dE \propto constant$  provides an equally good fit to the observed spectrum when it is corrected for absorption by only the GEBL (Dwek, Krennrich, \& Arendt 2005). Thus, without accurate knowledge of the intrinsic \gray\ spectrum it is impossible to confirm the extragalactic nature of the NIRBL. However, as shown below, it may be possible in specific cases to set limits on the nature of the NIRBL by examining the physical implications of the absorption-corrected spectrum. 

The blazar PKS2155--304 has been observed from over the entire optical to very high energy \gray\ wavelength regime. In particular it has been observed by the EGRET experiment on board the {\it Compton  Gamma Ray Observatory} (Vestrand, Stacy, \& Sreekumar 1995) at MeV--GeV energies, and most recently by the H.E.S.S. experiment at TeV energies (Aharonian et al. 2005a). Compared to H1426+428, this blazar has a more accurately determined TeV \gray\ spectrum, well approximated by a power law. Although not all the $\sim$ 10 eV to 3 TeV data is contemporaneous, the entire non-thermal spectral energy distribution (SED) of PKS2155--304 is well explained by the synchrotron self--Compton (SSC) model (Chiappetti et al. 1999) in which the blazar SED is characterized by a double peak: a synchrotron  peak located at UV to X--ray energies ($\sim$ 0.1 -- 1 keV), and a Compton peak located at energies of about 3 -- 300 GeV. The fit of the SSC model to the SED of PKS2155--304 spanning the eV to TeV region of its energy spectrum is presented in Figure 8 of Chiappetti et al. (1999). Dwek, Krennrich, \& Arendt (2005, Figure 7) have shown that if the observed PKS2155-304 spectrum is corrected for only GEBL absorption then the resulting SED is characterized by a smooth parabolic function with a peak around 10~GeV, consistent with the SSC model presented by Chiappetti et al. (1999). In contrast, corrected for the GEBL+NIRBL realization of the EBL the blazar's SED exhibits a very steep rise with $E^2dN/dE\propto E^{2.3}$ at TeV energies. 
In principle, it may be possible to reproduce a pileup of photons at TeV energies if these photons are produced by synchrotron emission from extremely high energy protons (Aharonian 2000). However, in this proton synchrotron model, even a steeply rising proton energy spectrum will produce an intrinsic \gray\ spectrum that is only mildly increasing with energy (see, for example, Figure 10 in Aharonian 2000). A pileup in the TeV spectrum of blazars can also be produced by the comptonization of ambient optical radiation by an ultrarelativistic cold jet with a bulk motion Lorentz factor of $\sim$ 10$^6$ - 10$^7$ emanating from the blazar (Aharonian, Timokhin, \& Plyasheshnikov 2002). While such scenario cannot be excluded,  it is not likely to to be the origin of the TeV emission from this blazar.  The observed spectrum of this blazar is very well represented by a power law, and it will require a very fine tuning of its intrinsic spectrum to produce this power law with the additional TeV opacity generated by the NIRBL.  

Alternatively, since the MeV--GeV EGRET and TeV H.E.S.S. data are not contemporaneous, it is possible that the SED has evolved in the time span between the observations so that  the GEBL+NIRBL corrected spectrum can be described by a simple SSC model, with a Compton peak at energies $\gtrsim$ 2~TeV. However, as pointed out by Dwek, Krennrich, \& Arendt (2005), a shift in the Compton peak by about two orders of magnitude would require a similar shift in the synchrotron peak energy, which is not supported by repeated X-ray observations of this blazar.

The conclusions of Dwek, Krennrich, \& Arendt (2005) are corroborated by  the recent detection of the most distant blazars H2356--309 ($z$ = 0.165) and 1ES1101--232 ($z$ = 0.186) (Aharonian et al. 2005b). Assuming  that intrinsic blazar spectra steeper than $E^2 dN/dE \propto E^{0.5}$ are physically unrealistic, Aharonian et al. (2005b) used these blazars to derive the strongest constraints on the EBL in the 0.8 -- 4 \mum\ wavelength region. The intensity they found is only slightly larger than the GEBL, leaving very little room for the contribution of Pop~III stars. 

%==================================
 \section{SUMMARY AND DISCUSSION}
 %==================================
 
We have shown that attributing the entire NIRBL to the energy output from Pop~III stars, as suggested by Salvaterra \& Ferrara (2003a), leads to several difficulties: 
\begin{enumerate}
\item it requires a peak star formation rate of {\it at least}  $\sim$ 1~\Msun\ \yr\ \Mpc, and more likely  about 2.5~\Msun\ \yr\ \Mpc, a value which is higher by a factor of 4 to 10 over that predicted from halo collapse models,
\item about 10\% of all baryons in the universe must be converted into Pop~III stars;
\item it predicts a \gray\ opacity leading to physically unrealistic absorption-corrected spectra of distant TeV blazars. 
\end{enumerate}
These difficulties suggest that at most, only a small fraction of the NIRBL can be of extragalactic origin. 

The spectrum of the NIRBL is almost identical to that of the zodiacal dust cloud. Therefore, the most plausible explanation for its origin is that it is reflected sunlight from the interplanetary dust cloud that, because of its isotropic spatial distribution, escaped detection by the DIRBE. However, such component may also contribute to the thermal infrared emission at $\lambda \gtrsim$ 10 \mum, requiring a reassessment of the EBL detections at far-infrared wavelengths. 
Furthermore, spatial fluctuations in the NIRBL cannot be attributed to the zodiacal dust cloud, although regarding the many uncertainties in its detailed spatial structure, such possibility cannot be entirely excluded.

Finally, we cannot exclude the possibility that the contribution of galaxies to the EBL may have been underestimated, and that the EBL intensity at optical/near-IR wavelengths is significantly larger than the extrapolations of Totani et al. (2001). Deeper surveys with the {\it Spitzer} Infrared Array Camera will be able to provide improved data on  the contribution of resolved galaxies to the EBL.  

Alternatively, a completely different source of energy, gravitational rather than nuclear, may be the origin of the NIRBL excess (Bond, Carr, \& Hogan 1986, Madau \& Silk 2005). Further 0.1 to 10 \mum\ measurements of the absolute brightness of the sky and of the zodiacal dust cloud will be crucial for  determining the true spectrum of the NIRBL, i.e.  the magnitude and location of the redshifted Ly$\alpha$ break in the EBL spectrum, and for resolving the origin of the NIRBL. Experiments such as the Cosmic Infrared Background Experiment (CIBER, PI: J. Bock), with the objectives of measuring the EBL spectrum around 1~\mum, and its spatial fluctuations in the I(0.90~\mum) and H(1.65~\mum) bands represent the next step in gaining further insight into the nature and origin of the NIRBL.
 
 {\bf Acknowledgements}
 We thank Fr\'ederic Galliano for running the CLOUDY code, and an anonymous referee for constructive suggestions that led to substantial improvements in the manuscript. ED acknowledges the support of NASA's LTSA 2004.

\clearpage

  %==================================
 %    BIBLIOGRAPHY
 %==================================

%  <=================== to use BIBTeX
% \bibliographystyle{aa}
 %\bibliography{/Users/edwek/science/science_bibliography}

%###### to use AASTex bibliography
% \input{/Users/edwek/science/bibliography_files/apj_bibliography}

 %==== table 1
 \begin{deluxetable}{lcc}
\tablecaption{Stellar and Nebular Energetics\tablenotemark{a}}
\tablewidth{0pt}
\tablehead{
\colhead{Emission component} &
  \colhead{ Luminosity (L$_{\odot}$)} &
    \colhead{ Temperature (K)}
   }
  \startdata 
  \sidehead{{\bf Input radiation} }
  stellar & 3.3 10$^7$ & 1.0 10$^5$ 	\nl
    \sidehead{\bf{Escaping radiation}}
    (1) stellar & 3.48 10$^6$ & \nodata \nl
    (2) nebular (cont.  + free-free) & 9.37 10$^6$ & 2.2 10$^4$ \nl
    (3) nebular lines & & \nl
    $\qquad$ Ly$\alpha$ (0.122 $\mu$m) 		& 1.68 10$^7$ & \nodata \nl
     $\qquad$ H$\alpha$ (0.656 $\mu$m) 		& 1.02 10$^6$ & \nodata \nl
   $\qquad$ H$\beta$  (0.486 $\mu$m)		& 3.38 10$^5$ & \nodata \nl
    $\qquad$ H$\gamma$  (0.434 $\mu$m)	& 1.58 10$^5$ & \nodata \nl
    $\qquad$  H$\delta$  (0.410 $\mu$m) 		& 8.33 10$^4$ & \nodata \nl
    $\qquad$   H$\epsilon$  (0.397 $\mu$m) 	& 5.15 10$^4$ & \nodata \nl
  \enddata
   \tablenotetext{a}{The stellar luminosity was calculated for a $M_*$ = 10$^3$~\Msun\ star radiating at the Eddington luminosity. Nebular energetics were calculated by CLOUDY for a nebula with primordial composition and a hydrogen number density of 10$^4$~cm$^{-3}$.  The total mass of the ionization bounded H~II region is 1.4$\times 10^3$~\Msun}
 \end{deluxetable}

%---- table 2 
  \begin{deluxetable}{lllll}
\tablecaption{Model Output Parameters\tablenotemark{a}}
\tablewidth{0pt}
\tablehead{
\colhead{Emission component} &
  \colhead{ } &
  \colhead{ NIRBL7} &
  \colhead{ } &
    \colhead{NIRBL9}
   }
  \startdata 
$n_0$ (\Mpc)	&  &7845 &   & 8301 \nl
$\Delta t$ (Myr) & &309&    & 238 \nl
$\dot \rho_*$ (\Msun\ \yr\ \Mpc) & &2.45 & & 2.59 \nl
$ \rho_0$ (\Msun\ \ \Mpc) & &$7.57\times 10^8$&    & $6.17\times 10^8$ \nl
$ \rho_*$/$\rho_b$ & &0.12&    & 0.10 \nl
$j_0$ (\Lsun\ \Mpc) &  &$2.60\times 10^{11}$ &  &    $2.74\times 10^{11}$ \nl
$\epsilon_*$ (erg\ \Mpc) &  &$9.7\times 10^{60}$&    & $7.9\times 10^{60}$ \nl
$\tilde E_b$ (erg/baryon) &  &$1.33\times 10^{-6}$&    & $1.08\times 10^{-6}$ \nl
\ \ \ \ (keV/baryon)& &  820&    & 670 \nl
$ I_{_{NIRBL}}$(\nwats)\tablenotemark{b} & &  74 &    &  46 \nl
  \enddata
   \tablenotetext{a}{Quantities were calculated for a $\Lambda$CDM model characterized by  $\Omega_{\Lambda}$ = 0.73,   $\Omega_m$ = 0.27, $\Omega_b$ = 0.044, and a Hubble constant of $H_0$ = 70~km~s$^{-1}$. For these parameters, the current baryonic mass and number densities are $\rho_b = 6.0\times 10^9$~\Msun\ \Mpc and $n_b = 7.35\times 10^{66}$ \Mpc. The effective temperature $\Delta t$ is defined in eq. (8).}
   \tablenotetext{b}{$ I_{_{NIRBL}}$ represents the contribution of Pop~III stars to the EBL  intensity, integrated over all wavelengths. Integration over only the $\sim$ 1.4 - 4.0~\mum\ wavelength interval will give an value of only $\sim$ 30 \nwats. This integration will omit the lowest redshift Ly$alpha$ emission (primarily for model NIRBL7) and all of the redshifted Balmer emisison.}
 \end{deluxetable}

\end{document}